\documentclass[aps,pra,preprint]{revtex4}
\usepackage{graphicx}
\usepackage{latexsym}
\usepackage{amssymb}
\usepackage{epsfig}
\begin{document}
\hspace {-8mm} {\bf \LARGE Exchange Potential for excited states: A selfconsistent DFT calculation.} \\ \\

\hspace {-8mm} {\bf Md.Shamim and Manoj K. Harbola}\\
{Department of Physics, Indian Institute of Technology,
Kanpur 208016, India}

\begin {abstract}

An LDA exchange potential is proposed for excited states and to test this potential we apply it to the excited states of atomic system. The potential is an 
approximate functional derivative of an accurate exchange energy functional for excited states. We show that the potential satisfies Levey-Perdew theorem for exchange energy
and janak theorem for orbital energy very well in excited state cases.  The potential is the first of its kind for excited state and can easily be generalized 
to other excited states of interest. We compare our results  with those of other approaches reported in the literature.
\end {abstract}

\maketitle
\section {Introduction}

Density Functional Theory (DFT) as founded on the works of Hohenberg and Kohn \cite {hk} and Kohn and Sham \cite {ks} is very successful for the ground states \cite {parryang}.
It states that total energy of an interacting many particle system can be written as the functional of density of the system.  ie .

\begin {equation}
E=E \left[\rho \right]
\label {energy1}
\end {equation}

The Kohn Sham version of the theory reduces the many interacting-particle problem to a virtual non-interacting many particle problem.  Therefore
total energy E is written explicitly in terms of contributions from kinetic and potential part as,

\begin {equation}
E\left [ \rho \right ] = T_{s}\left [ \rho \right ] + \int \rho ({\bf r})v_{ext}({\bf r})d{\bf r} + E_{XC}\left [ \rho \right ]
\label {energy2}
\end {equation}

Where $ T_{s}\left [ \rho \right ]$ is the non-interacting kinetic energy of the system, $ v_{ext}$ is the external potential and $ E_{XC}\left [ \rho \right ]$ is the exchange-correlation energy functional. In atomic system nuclear potential acts as the  external potential for the electrons.  The density is obtained by solving 
self-consistently a set of single particle Kohn-Sham equations for virtual non-interacting system.

\begin{equation}
\left[-\frac{{\nabla}^{2}}{2}+ \int\frac{\rho({\bf r'})}{|{\bf r}-{\bf r'}|}d{\bf r'}+v_{ext}({\bf r})+v_{xc}({\bf r}) \right ] \phi_{i}({\bf r})
=\epsilon_{i}\phi_{i}({\bf r})
\label{kseq}
\end{equation}

Where, $v_{xc}({\bf r})$ is the exchange correlation potential and is defined as,

\begin{equation}
v_{xc}(\bf r)= \frac{\delta E_{xc}(\rho)}{\delta\rho}
\label{xcp}
\end{equation}

Usually the exchange-correlation potential is split into exchange and correlation potential. ie
.$ v_{xc}(\left [ \rho \right ];\bf r) = v_{x}(\left [ \rho \right ];\bf r) +  v_{c}(\left [ \rho \right ];\bf r)$.    
 The maping from an interacting system to a non-interacting system is exact but none of the $v_{x} $ and $v_{c} $ is known exactly in a form 
that can be used in the calculations for the practicle purposes.  Therefore  we need to use approximation for the exchange and correlation potentials.  
Widely used approximation in DFT is the local density approximation(LDA) for the exchange and correlation part.

The success of ground state DFT rests on the existence of accurate LDA functionals for exchange energy, correlation energy and also on the existence of corresponding  
potentials. For exchange potential most often Dirac's exchange potential for homogeneous electron gas \cite{Dirac} is used in LDA sense.  Further 
depending upon the requirements asymtotic corrections \cite{LB} or gradient corrections \cite {perdvos} are added to the Dirac exchange potential to get better results.  For 
ground states almost accuracy of chemical interest has been achieved \cite{affinity} with DFT.  The theory has been applied to study the ground state properties of
finite as well as extended system and quite resonable predictions could be made \cite{kamal}.  Other aprroximations for exchange potential to mention are 
Slater's averaged potential \cite{slater} , Harbola-Sahni Fermi-hole potential \cite {harbola1}, a simple effective exchange potential by Becke et al \cite {becke}, 
optimized effective exchange potentials \cite{oep1,oep2}, Hartree-Fock exchange potential\cite {hartree}.  For finite systems contribution from exchange part is much
 larger than that from the correlation part, therefore our work in this paper is concerned about the former. 

In DFT it has been a general practice to develop accurate exchange energy functional first, and to calculate exchange potential functional derivative of the exchange energy 
functional is taken with respect to total density. Another way could be like we construct exchange potential first, by using some 
physical arguments and to calculate corresponding exchange energy we can either use the exchange energy functional reported by us in our earlier work \cite{samalh,sami} 
or we can use the Levy-Perdew relation for exchange energy. For ground state LDA both approaches lead to the same result. Whereas, for the excited 
states LP relation leads to inacurate results if the excited state potential is not a very good approximation.  

Physics and Chemistry are full of examples where studies of the excited states of  many systems is an active area of research \cite {kamal} and DFT can always be 
applied to most of such studies.  Extension of density functional formulation to excited states is now well stablished \cite {levynagy,gor,samal2}.  However similar to 
the ground state case the implementation of DFT to excited requires accurate exchange and correlation energy functionals and corresponding accurate potentials. 

A genuine first step towards excited states started with the application Dirac exchange potential itself to the excited states.  From the studies of 
Gunnarson and Lundqvist \cite {gunnar} and von Barth \cite {von} it is known that with ground state functionals only energies of the lowest states of each 
symmetry can be determined.  The use of ground state LDA potential for excited states works well for some cases but fails miserably for many.\cite{samalh}.
  The reason is that the only way ground states functionals incorporate the symmetry of the system is through the density. Till now most of the  
calculations in both time dependent and time independent versions of DFT has employed the ground state LDA exchange potential.  Despite its 
 limitations the ground state LDA potential is in widespread use for excited calculations.  The reason is that, it is an orbital independent potential while, 
for excited states the exchange potentials become  orbital dependent.  Therefore Kohn Sham type calculation is not possible with such type of orbital dependent potentials.  
Due to this limitation the ground state LDA potential becomes an obvious choice . 

There has been some intermittent attempt to construct accurate exchange potential for excited states.  Gaspar \cite {gasp} and Nagy's\cite {nagyvx} works in this
direction are a few to mention.  They have given an ensemble averaged exchange potential for the excited states and have used their potential
to caluclate excitation energy for single electron excitations. However in such calculations the beauty of ground state like density functional calculation for
individual excited state is always missing.  Therefore, we attempt to develop an LDA excited state exchange potential for the excited states
so that density functional calculation could easily be done for individual excited states in as straight forward way as for the ground states.

In this paper we report the construction of an exchange potential for excited states.  This exchange potential is orbital dependent but for different
 classes of excited states it becomes  independent of the orbitals.  For example, in the case of Ne with 2s electrons excited to the 3s orbitals, exchange potential we 
report here will be same for all electrons but if we consider a different class of excitation in which 2 electrons from 2p orbitals are excited to 3s, again exchange
potential will be same for all the electrons but it will be different from that of the previous case.  Therefore  Kohn-Sham type calculations can easiy be done for the 
excited states using the potential reported here.  The idea can be generalized to any case of interest and is easy to implement in
 doing self consistent field calculations. 

The potential is basically a generalization of Dirac exchange potential for excited states by using split k-space. The conceptual motivation to construct 
such potential comes from the the fact that the ground state Hartree-Fock exchange potential for the highest occupied orbital(HOMO) equals the Functional 
derivative of exchange energy with respect to denstiy. ie.
\begin {equation}
v_{i}^{HF}({\bf r})|_{i=HOMO} = \frac{\delta E_{x}^{LDA}({\rho})}{\delta{\rho({\bf r})}}
\label {gxp}
\end {equation}

Therefore, for the excited state we do the same.  We calculate the Hartree-Fock exchange potential which is orbital dependent and take the potential 
for each electron to be equal to the potential corresponding to the upper most orbital(HOMO).  In the following we show that this 
is equal to the functional derivative of modified exchange energy functional reported by us\cite{samalh} for excited states with respect to 
the density corresponding to the largest wave-vector in the k-space.  We have been persuing the idea of constructing the 
energy functionals for excited-states by using split k-space for the past few years with significant success. We have shown that accurate exchange 
energy and kinetic energy functional can be constructed in this manner \cite {samalh,hem}.  In the present work we have employed the same idea to construct the 
exchange potential for excited states so that excited state calculation could be performed with much simplicity.  Work on correlation energy functional 
will be presented elsewhere.

The outline of the paper is as follows. In the sub-section II-(a) we describe the construction of potential as derived on the idea taken from the ground state that
HF potential for HOMO becomes the exchange potential for all the electrons of the system. In sub-sction II-(b) construction of exchange potential from the 
functional derivative of the exchange energy functional is described for the excited states. In sub-sction II-(b) we also describe how we change the exchange potential to make
it better. Results are presented in section  III and  we conclude in section IV.

\begin{figure}[thb]
\includegraphics[width=5.5in,angle=0.0]{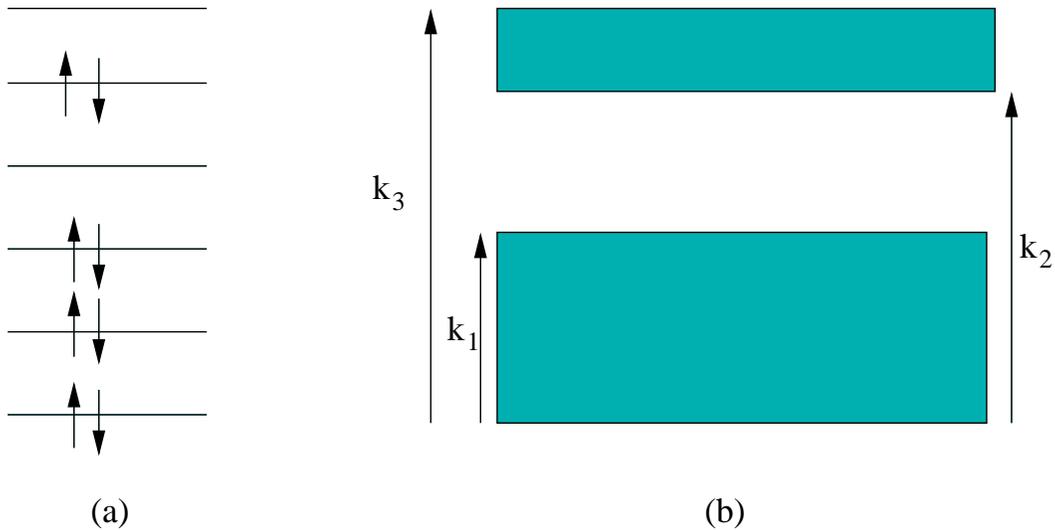}
\caption{Orbital and the corresponding $k-space$ occupation in the
excited state configuration of a homogeneous electron gas(HEG).}
\label{k-space}
\end{figure}

\section {Construction of Exchange potential for excited-states}
To construct an LDA exchange potential for excited states we map the excited state density to corresponding k-space for homogeneous electron 
gas(HEG) as shown in figure 1. Unlike ground state case now k-space has some gap(s) corresponding to the missing orbitals.  The  exchange potential for an excited state of  HEG
can be obtained in two ways.(i) From the Hartree-Fock expression for exchange potential and (ii) From the functional derivative of  exchange energy of the electrons
taken with respect to total density. In the following we describe the two methods one by one.

Here, it should be noted that although the potential constructed in this way depeneds on the densities $\rho_{1}$, $\rho_{2}$, $\rho_{3}$ corresponding to the 
wave vectors $k_{1}$,  $k_{2}$, $k_{3}$, the functionals obtained from this potential intrinsically depend on the ground state density and excited state density.
 Therefore this approach of doing excited state studies in DFT is in conformity with the excited state formulation of DFT by Levy, Nagy \cite {levynagy} and Gorling \cite {gor} 
and also by our group \cite {samal2,samalth}.

\subsection {LDA Exchange potential from Hartree-Fock exchange potential}

The Hartree-Fock exchange potential for a system of fermions is given by

\begin {equation}
v_{x}^{i}=v_{x}(\phi_{i})=-\sum_{j}\int\frac{\phi^{\ast}_{j}({\bf r'})\phi^{\ast}_{i}({\bf r'})\phi_{j}({\bf r})}
{\phi_{i}({\bf r})|{\bf r}- {\bf r'}|}d{\bf r'}
\label {hfx}.
\end {equation}

for homogeneous electron gas,
\begin {equation}
\phi_{{\bf k}_{i}}({\bf r}) = \frac{1}{{\sqrt V}} e^{ \left (i{\bf k}_{i}\cdot {\bf r}\right )} 
\label {wf}.
\end {equation}

Using this form of wavefunction in equation \ref {hfx} we get an exchage potential for one-gap systems shown in figure 1 to be given by,

\begin {equation}
v_{x}^{i}=\frac {1}{\pi} \left [-k_{1}+k_{2}-k_{3} - \frac {k_{1}^{2}-k_{i}^{2}}{2k_{i}}\ln\left|\frac{k_{i}+k_{1}}{k_{i}-k_{1}}\right|
+ \frac {k_{2}^{2}-k_{i}^{2}}{2k_{i}}\ln\left|\frac{k_{i}+k_{2}}{k_{i}-k_{3}}\right|-\frac {k_{3}^{2}-k_{i}^{2}}{2k_{i}}\ln\left|\frac{k_{i}+k_{3}}{k_{i}-x_{3}}\right|\right ]
\label {hfxpi}
\end {equation}

Which is  an orbital dependent potential. To make this potential an orbital independent potential we draw the analogy from the ground state exchange potential, where 
the exact LDA potential happens to be equal to the HF potential for HOMO. Therefore we take the potential seen by the electron in HOMO as the exchange potential  
for all the electrons.  For this we put $k_{i}=k_{3}$ , with this the Hartree-Fock exchange potential of eqn \ref {hfxpi} for the excited state reduces to 
the following expression for exchange potential .
\begin {equation}
v_{x}^{MLSDA}=\frac{k_{3}}{\pi}\left [-1+x_{2}-x_{1}+\frac{1}{2}(1-x_{1}^{2})\ln\left|\frac{1+x_{1}}{1-x_{1}}\right|
-\frac{1}{2}(1-x_{2}^{2})\ln\left|\frac{1+x_{2}}{1-x_{2}}\right|\right ]
\label {xphf}
\end {equation}
where,
\begin {equation}
x_{1}=\frac{k_{1}}{k_{3}}, x_{2}=\frac{k_{2}}{k_{3}}
\label {xs}
\end {equation}

 The wave-vectors ${\bf k} $'s  are shown in figure 1 and are related to the densities of core electrons $\rho_{core} $,
missing electrons$ \rho_{missing}$ and shell electrons $\rho_{shell}$ as give below

\begin {equation}
k_{1}^{3}=3\pi^{2}\rho_{core}
\label {k1}
\end {equation}

\begin {equation}
k_{2}^{3}-k_{1}^{3}=3\pi^{2}\rho_{missing}
\label {k2}
\end {equation}

\begin {equation}
k_{3}^{3}-k_{2}^{3}=3\pi^{2}\rho_{shell}
\label {k3}
\end {equation}

\subsection {LDA Exchange potential from functional derivative}

If we know $E_{X}(\rho)$ then from its functional derivative we can get exchange potential. But the exact exchange energy functional 
is not known.  Therefore we need to first have an accurate approximation for exchange energy of HEG in excited state.  In our previous work \cite{samalh} we have shown 
that for all systems having configuration same as that shown in figure 1,the LDA approximation to the exchange energy functional is given by 

\begin{eqnarray}
E_{X}^{MLSDA} &=& \int\rho({\bf r})\left[\epsilon(k_{3}) - \epsilon(k_{2})+ \epsilon(k_{1})\right]
d{\bf r}
+ \frac{1}{8\pi^{3}}\int(k_{3}^{2}-k_{1}^{2})^{2}\;ln\left(\frac{k_{3}+k_{1}}{k_{3}-k_{1}}
\right)d{\bf r} \nonumber \\
&-&\frac{1}{8\pi^{3}}\int(k_{3}^{2}-k_{2}^{2})^{2}\;ln\left(\frac{k_{3}+k_{2}}{k_{3}-k_{2}}\right)
d{\bf r}  
-\frac{1}{8\pi^{3}}\int(k_{2}^{2}-k_{1}^{2})^{2}\;ln\left(\frac{k_{2}+k_{1}}{k_{2}-k_{1}}\right)
d{\bf r}
\label{eq:xmlda}
\end{eqnarray}

Where MLSDA stands for modified local spin density approximation. Here $\epsilon(k_{i})=-\frac{3k_{i}}{4\pi}$ is the exchange energy per electron for the HEG in its ground-state
with the Fermi wave-vector equal to $k_{i}$.

The exchange energy functional given by equation \ref{eq:xmlda} is a highly accurate approximation for excited states of one gap systems.  This has been proved by 
calculation of accurate excitation energies employing it \cite {samalh} also by band gap calculations \cite {rahman} .  Therefore if we take functional 
derivative of this functional  with respect to density we should get an accurate modified LDA exchange potential for this class of excited states. i.e.

\begin{equation}
v_{x}^{MLSDA}(\bf r)= \frac{\delta E_{X}^{MLSDA}(\rho)}{\delta\rho}
\label{xp}
\end{equation}
 
It is not possible to get a workable analytical expression for $ v_{x}^{MLSDA}({\bf r})$ out of eqn \ref{xp}.  Therefore on the basis physical argument
that the chemical potential of the system  corresponds to the highest occupied orbital, we instead take the functional derivative of exchange energy 
functional with respect to $\rho_{3}$, which is the density corresponding to the highest wave vector $k_{3}$ .  Therefore we now assume the exchange potential 
to be defined as,
\begin{equation}
v_{x}^{MLSDA}(\bf r)= \frac{\delta E_{x}^{MLSDA}(\rho)}{\delta\rho_{3}}
\label{xp3}
\end{equation}

where,
\begin {equation}
\rho_{3}({\bf r}) = \frac {k_{3}^{3}}{3\pi^{2}}
\label {rho3}
\end {equation}

If we take functional derivative in this way the potential obtained is excacly equal to that given by equation \ref{xphf}.

We have reached the same result from two different physical arguments, this in some sense assures us about the correctness of the approach taken.  
Therefore we expect this potential to work well.

The exchange potential derived above satisfies LP theorem and Janak theorem quite well. The excitation energies obtained with this 
potential with exchange energy calculated using MLSDA fucntional are very accurate.  However if instead of the MLSDA fucntional we use LP relation to calculate exchange 
energy then excitation energies calculated in this way are good only for few cases. In many cases excitation energies deviate from the exact values more than the LSD values
.Therefore, we studied various cases and based on our analysis we suggest a change in the potential of  equation \ref {xphf} . 
We propose the new form of the potential to have an orbital dependent exponent so that it will work well for all types of excitation. 
With such change the potential takes the following form.

\begin {equation}
v_{x}^{M}=\frac{k_{3}}{\pi}\left [-1+x_{2}-x_{1}+\frac{1}{2}(1-x_{1}^{a^{\sigma}})\ln\left|\frac{1+x_{1}}{1-x_{1}}\right|
-\frac{1}{2}(1-x_{2}^{a^{\sigma}})\ln\left|\frac{1+x_{2}}{1-x_{2}}\right|\right ]
\label {xphfx}
\end {equation}

Where the exponent $a^{\sigma}$ depends upon the number of electrons missing from an orbital and on the degeneracy of that orbital and its value
is determined by using the formula
\begin {equation}
a^{\sigma} = \frac {2N_{m}^{\sigma}}{d^{\sigma}+1}
\label {asig}
\end {equation}
where, $N_{m}^{\sigma}$ and $d^{\sigma}$ are the number of missing electrons and the degeneracy of the orbital from which the electrons are
 missing respectively and $\sigma $ is the spin index of the electrons.  Superscript M on exchange potential stands for modified or say modelled.

The new modified exchange potential for excited states as given by equation \ref {xphfx} improves upon the LDA results uniformly for almost all cases to which
it was applied. This potential has similar behaviour as the LDA potential, Therefore the asymptotic corrections need to be invoked from out side  and 
further to achieve the chemical accuracy the gradient corrections will also be needed. But over all  the potential reduces the LDA error for excited states substantially. 

Further, the exchange potential given by equation \ref {xphfx} is independent of orbitals for one gap system,therefore using it a self-consistent
Kohn-Sham calculation can be done very systematically. 

In doing self-consistent calculations we do not include self-interaction correction (SIC) term in the exchange potential.  However for calculating excitation 
energy we do incorporate SIC for those orbitals which take part in transition and create a gap \cite{samalh} . As mentioned above, we use Levy-Perdew relation 
to calculate exchange energy from the corresponding potential for excited states.  The exchange energy functional after SIC correction takes the following form,

\begin {equation}
E_{X}^{MSIC}=E_{X}^{M}-\sum_{i}^{rem}{E_i}^{SIC}-\sum_{i}^{add}{E_i}^{SIC}
\label{18}
\end {equation}
where,
\begin {equation}
E_{i}^{SIC}\left[\phi_i\right]= \int\int\frac{|\phi_{i}({\bf r}_{1})|^{2}|\phi_{i}
({\bf r}_{2})|^{2}}{|{\bf r}_{1}-{\bf r}_{2}|}d{\bf r}_{1}d{\bf r}_{2} 
+E^{LSD}_{X}\left[\rho\left[\phi_i\right]\right]
\label{19}
\end {equation}

and $E_{X}^{M}$ is obtained using $v_{x}^{M}$ potential in Levy-Perdew relation 
\begin {equation}
E_{x}^{M}=-\int\rho({\bf r}){\bf r}.\nabla v_{x}^{M}d{\bf r}
\label {LP}
\end {equation}

\section {Test for the Exchange potential}

The potential we are reporting here is not an exact functional derivative of the exchange energy functional.  Therefore, two genuine questions arise 
(i) How good this potential is,  and (ii) Is it better than the LDA potential for excited states?  To answer these two questions we test whether this potential
satisfies the following two theorems ,

a) Levy-Perdew theorem,

\begin {equation}
E_{X}=-\int\rho({\bf r}){\bf r}.\nabla v_{x}d{\bf r}
\label {LP}
\end {equation}

Where $\rho ({\bf r})$ is the total density and in place of $v_{x}$ we use potential of equation \ref {xphfx}.

b) Janak Theorem

\begin {equation}
\frac {\partial E}{\partial n_{i}}=\epsilon _{i}
\label {JNK}
\end {equation}

Where $\epsilon_{i}$ is the orbital energy and $n_{i}$ is the occupation of that orbital and $E$ is the total energy.

Theorem (a) is a severe test for exchange potentials.  We show in figure \ref{dexf1} and figure \ref{dexf2} that the Exchange energy obtained from the
 potential of equation \ref{xphfx} using Leve-Perdew theorem is much closer to those obtained with the Exchange energy functional of equation \ref{eq:xmlda} 
directly.  While the difference is much larger when LDA exchange potential is used to obtain exchange energy.  Therefore we can say that potential is a 
reasonably good and is better than the LDA potential for excited states. 

Next we calculate change in total energy of the system  by changing the occupation of the orbitals and then calculate the gradient of total energy$(E)$ with respect to 
occupation of orbital $n$. We show that this gradient is very close to the orbital energy of the corresponding orbital.In this way we test the Janak theorem.  We find that
the potential of equation \ref {xphfx} satisfies the Janak theorem very well.

\section {Results}

In this section we report the results obtained with the exchange potential constructed in the section II. We use Hermann Skillman program for our calculations 
with some changes to incorporate the new potential for excited states. All the calculations are fully self-consistent and as simple as the ground state calculations. 
Since LDA as well as all its modified forms are good for the states which can be represented by single determinants, therefore we do calculation for a particular configuration
rather than a state.  Further all our calculations are done in cnetral field approximation, that is we take density to be spherical . This approximation is justified becuase 
the non-sphericity of density does not make much difference\cite{janakw}.

The  results presented here are for the class of the systems  which have one gap in the occupation of orbitals.
 To clearly show the advantage of constructing exchange potential for excited states we compare all our results with the corresponding results obtained with ground 
state LSD exchange potential and for excitation energies we have compared our results with standard results also wherever we could do.

We first present the results for the test of Levy-Perdew theorem and Janak theorem in the sub-sections IV-A and IV-B respectively. The results peresented in these two sections
are for one gap cases. In sub-section IV-C we discuss the calculation of excitation energies for one-gap systems.

\subsection {Test for the Levy-Perdew theorem}
\begin{figure}[thb]
\includegraphics[width=4.5in,angle=-90.0]{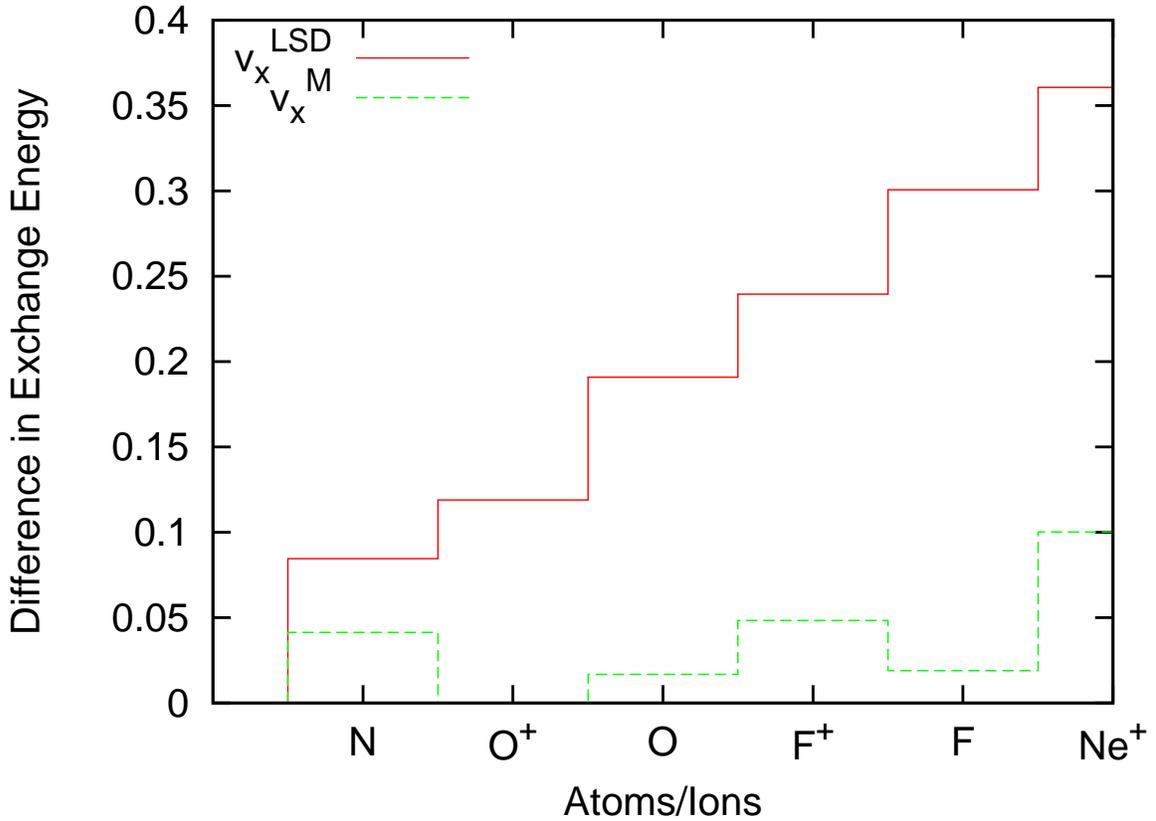}
\caption{ $\Delta E_{X} =\left |  E_{X}^{LP}-E_{X}^{MLSDA} \right | $ for $v_{x}^{M}$ and $v_{x}^{LSD}$ potentials for the atoms/ions shown on the horizontal axis}
\label{dexf1}
\end{figure}

\begin{figure}[thb]
\includegraphics[width=4.5in,angle=-90.0]{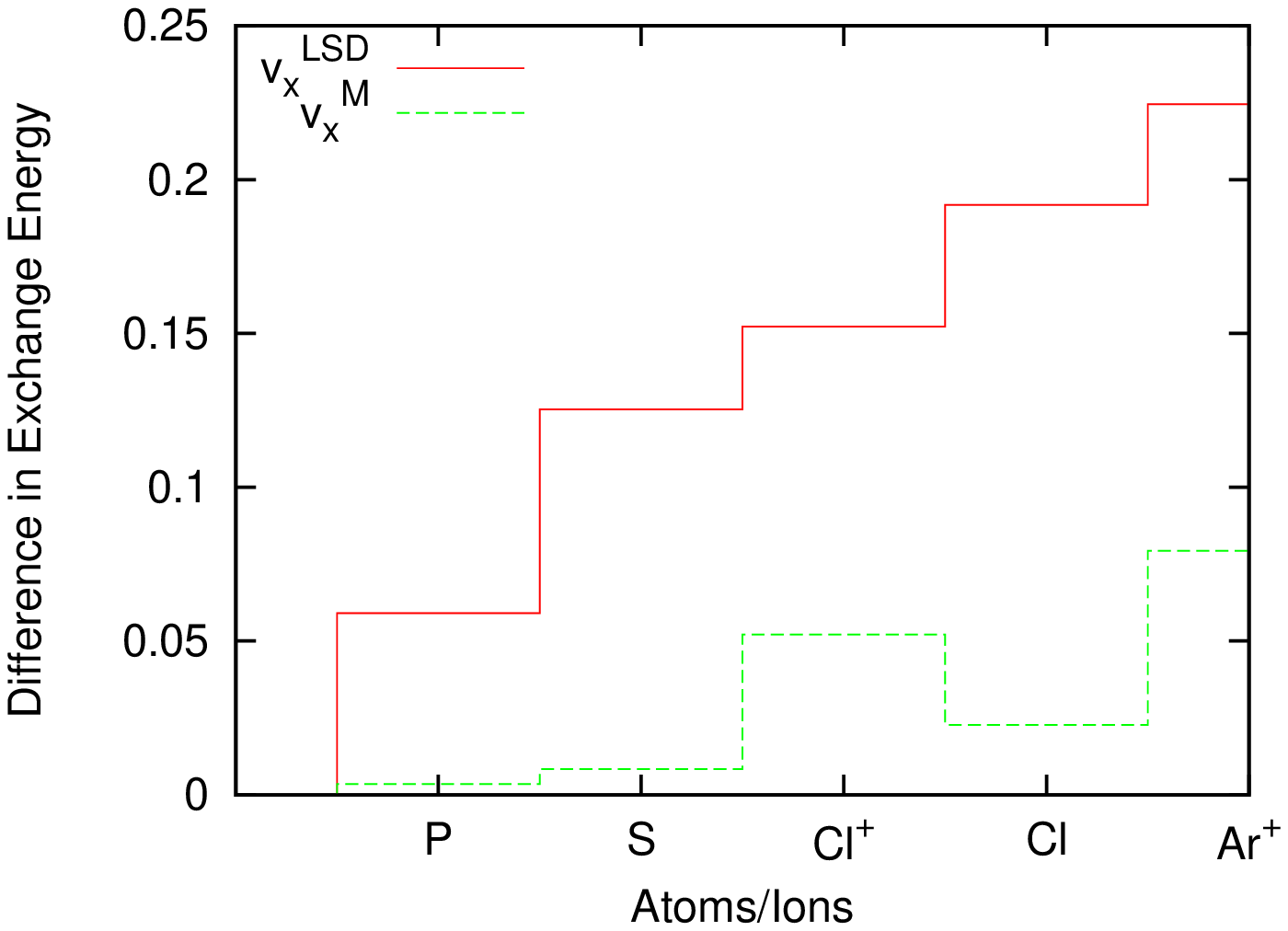}
\caption{ $\Delta E_{X} =\left |  E_{X}^{LP}-E_{X}^{MLSDA} \right | $ for $v_{x}^{M}$ and $v_{x}^{LSD}$ potentials for the atoms/ions shown on the horizontal axis}
\label{dexf2}
\end{figure}

We have calculated difference in exchange energy as calculated using Levy-Perdew relation \ref {LP} and the exchange energy functional \ref {eq:xmlda} with MLSDAX and LSD 
exchange potentials. We find that for most of the cases this difference is very small for the MLSDAX potential as compared to the LSD exchange potential. This shows that 
for excited states MLSDAX exchange potential is closer to the excact exchange potential than the LSD potential is. The results for the configurations of atoms and ions given
in table-I and table-III are displayed in figure \ref {dexf1} and figure \ref {dexf2} respectively.

\subsection {Test for Janak Theorem}
\begin{figure}[thb]
\includegraphics[width=4.5in,angle=-90.0]{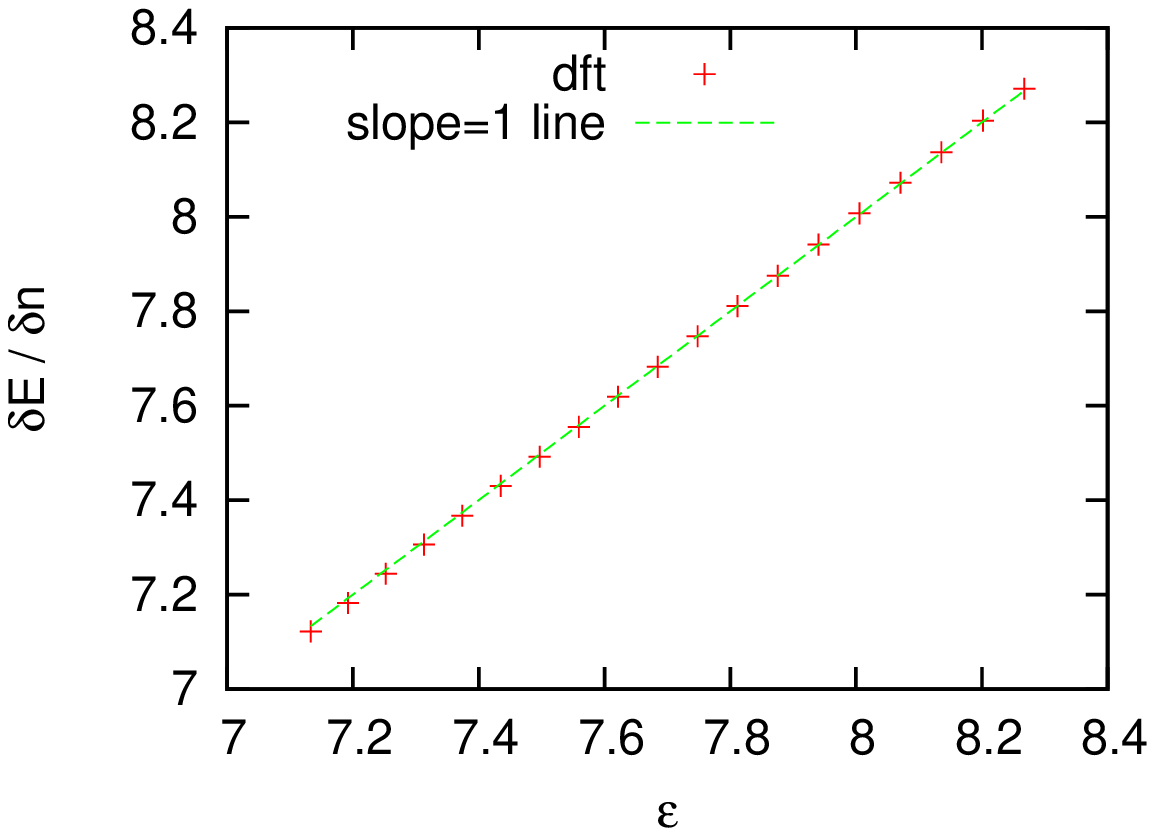}
\caption{Drivative of energy with respect to the orbital occupation versus the orbital energy and slope equals one line.}
\label{janak1}
\end{figure}

Change in total energy is sensitive to the nature of exchange potential. Therefore, if the modified exchagne potential reported here is accurate one, it should
satisfy Janak Theorem.  To test this we take a configuration of an atom with one gap and vary the occupation of a particular orbital. As the occupation is changed 
from given value to a value less by one the corresponding total calculated using the MLSDA functional and the orbital energy of that orbital is noted for each 
intermediate occupations. From the table generated in this way we calculate the slope of energy with respect to the occupation of the orbital. 
Then we plot the slope against the orbital energy. If Jank theorem is satisfied the curve obtained in this way should coincide with a line having slope 
equal to one. And this also tell us that to a very good degree the exchange potential is a functional derivative of the MLSDA functional.
 The result for Cl atom having configuration $1s^{2}2s^{2}2p^{6}3s^{1}3p^{6};M_{l}=0,M_{s}=1/2$ is shown in figure \ref {janak1}. Here orbital considered 
is the up spin $2p$ orbital.

\subsection {Transition energy calculation}
Now we employ the potential constructed in section-II to calculate transition energies for various configurations of atomic systems. 
Our calcualtion is $\Delta SCF$ type . For ground states excited-state potential reduces to regular ground state LSD exchange potential, 
Therfore we do two self-consitent calculations one for ground state and one for the excited state, and from the difference of the two 
we get the transition energies. When we use the MLSDA fucntional to calculate the exchange energy, the excitation energy obtained in this way are very accurate. This 
has been reported in our previous works \cite{samalh,sami}. Here we show that the excited state exchange potential leads to reasonably accurate 
excitation energies even when the exchange energy is calculated through the LP theorem using this potential.

In tables I-VIII we present transition energis for one-gap systems. The results given in the talbe I-VII are for excitations of one electron for different cases 
and in table-VIII transition energies for excitation of two electrons are given. The results are compared with HF,TDDFT results and we also give the excitation energies 
obtained by employing the MLSDA functional for the exchange energy.  We see that the excitation energies obtained with the potential derived above are comparable to those 
obtained by employing MLSDA functional for most of the cases. Only in two to three of all the cases studied here not very good and due to these few cases having 
slightly off numbers the average errors are somewhat larger in the corresponding tables.  The average percentage error in excitation energy calculated with LSD potential for
tables 1 to VIII are respectively 14.30, 2.80, 15.80, 5.86, 3.86, 12.05, 2.95, 16.14. In table-II we do not get good result for Li and in table V and VII there are two such cases
which raise the average errors. Except these, in all other cases the new potential gives either substantial improvement over LSD results or is as good as LSD is. 
  We are working on the potential to make it even better and workable without exception. 

The configurations of the excited taken here are exactly same as that in \cite{samalh} for one-gap systems.  Therefore more details about these can be found there.

\section{Conclusion}
We have proposed an LDA exchange potential for excited states. We show that the exchange potential reported here satisfies 
Levy-Perdew theorem and Janak therem very well and also it gives resonably good excitation eneregies for most of the cases studied 
here.  Although it is orbital dependent, for particular classes of excited it becomes same for all electrons,
 therefore ground state like K-S calculation can be performed easily.  We have employed this potential to one-gap systems with encouraging success.

\newpage

\begin{table}
\caption{Transition energies, in atomic units, of an electron being excited from the
$2s$ orbital of some atoms to their $2p$ orbital.  The first column gives this energy
as obtained in Hartree-Fock theory.  The numbers in the second column are obtained by
employing the ground-state LSD potential and MLSDA functional for excited-state. Third column 
gives transition obtained with potential of equation \ref{xphf} and in the fourth column are the numbers  
obtained with potential of equation \ref{xphfx}.  The last
column gives the transition energies obtained by employing TDDFT.}
\vspace{0.2in}
\begin{tabular}{lccccc}
\hline
atoms/ions & $\Delta$$E_{HF}$ & $\Delta$$E^{a}$ & $\Delta$$E^{b}$ & $\Delta$$E^{c}$ &
$\Delta$$E_{TDDFT}$  \\
\hline
$N(2s^{2}2p^{3}\;^{4}S\rightarrow2s^{1}2p^{4}\;^{4}P)$ &0.4127&0.4014&0.4954&0.4549&0.4153 \\
$O^{+}(2s^{2}2p^{3}\;^{4}S\rightarrow2s^{1}2p^{4}\;^{4}P)$ &0.5530&0.5571&0.6452&0.5581&0.5694 \\
$O(2s^{2}2p^{4}\;^{3}P\rightarrow2s^{1}2p^{5}\;^{3}P)$ &0.6255&0.6214&0.6821&0.6460&0.5912 \\
$F^{+}(2s^{2}2p^{4}\;^{3}P\rightarrow2s^{1}2p^{5}\;^{3}P)$ &0.7988&0.8005&0.8399&0.7540&0.7651 \\
$F(2s^{2}2p^{5}\;^{2}P\rightarrow2s^{1}2p^{6}\;^{2}S)$ &0.8781&0.8573&0.8651&0.8445&0.7659 \\
$Ne^{+}(2s^{2}2p^{5}\;^{2}P\rightarrow2s^{1}2p^{6}\;^{2}S)$ &1.0830&1.0607&1.0348&0.9620&0.9546 \\
\hline
Average error &  & 1.78\%& 9.45\% & 5.70\%& 
\end{tabular}
\end{table}

\footnotetext[1]{a With MLSDA functional of equation \ref{eq:xmlda}}
\footnotetext[2]{b With the excited-state potential of equation \ref{xphf} }
\footnotetext[3]{c With the modified excited-state potential of equation \ref{xphfx} }

\begin{table}
\caption{The caption is the same as that for Table 2 except that we are now
considering transitions from the outermost orbital to an upper orbital for
weakly bound systems.}
\vspace{0.2in}
\begin{tabular}{lccccc}
\hline
atoms/ions & $\Delta$$E_{HF}$ & $\Delta$$E^{a}$ & $\Delta$$E^{b}$ & $\Delta$$E^{c}$ &
$\Delta$$E_{TDDFT}$  \\
\hline
$Li(2s^{1}\;^{2}S\rightarrow2p^{1}\;^{2}P)$ &0.0677&0.0672&0.0973&0.09217&0.0724 \\
$Na(3s^{1}\;^{2}S\rightarrow3p^{1}\;^{2}P)$ &0.0725&0.0753&0.0669&0.0774&0.0791 \\
$Mg^{+}(3s^{1}\;^{2}S\rightarrow3p^{1}\;^{2}P)$ &0.1578&0.1696&0.1800&0.1580&0.1734 \\
\hline
Average error &  & 6.30\%& 21.60\% & 14.30\%& 
\end{tabular}
\end{table}

\begin{table}
\caption{Electron transition energy from the $3s$ to the $3p$ orbital in some atoms.}
\vspace{0.2in}
\begin{tabular}{lccccc}
\hline
atoms/ions & $\Delta$$E_{HF}$ & $\Delta$$E^{a}$ & $\Delta$$E^{b}$ & $\Delta$$E^{c}$ &
$\Delta$$E_{TDDFT}$  \\
\hline
$P(3s^{2}3p^{3}\;^{4}S\rightarrow3s^{1}3p^{4}\;^{4}P)$ &0.3023&0.3055&0.3380&0.3027&0.3183 \\
$S(3s^{2}3p^{4}\;^{3}P\rightarrow3s^{1}3p^{5}\;^{3}P)$ &0.4264&0.4334&0.4877&0.4246&0.4122 \\
$Cl^{+}(3s^{2}3p^{4}\;^{3}P\rightarrow3s^{1}3p^{5}\;^{3}P)$ &0.5264&0.5403&0.5983&0.5419&0.5113 \\
$Cl(3s^{2}3p^{5}\;^{2}P\rightarrow3s^{1}3p^{6}\;^{2}S)$ &0.5653&0.5630&0.6181&0.5400&0.4996 \\
$Ar^{+}(3s^{2}3p^{5}\;^{2}P\rightarrow3s^{1}3p^{6}\;^{2}S)$ &0.6766&0.5174&0.7264&0.5965&0.6007 \\
\hline
Average error &  & 1.15\%& 11.28\% &3.92\% & 
\end{tabular}
\end{table}

\begin{table}
\caption{Electron transition energy from the $2s$ to the $3p$ orbital in the
same atoms as in Table 4.}
\vspace{0.2in}
\begin{tabular}{lccccc}
\hline
atoms/ions & $\Delta$$E_{HF}$ & $\Delta$$E^{a}$ & $\Delta$$E^{b}$ & $\Delta$$E^{c}$ &
$\Delta$$E_{TDDFT}$  \\
\hline
$P(2s^{2}3p^{3}\;^{4}S\rightarrow2s^{1}3p^{4}\;^{4}P)$ &6.8820&6.9564&6.8165&6.6290&6.1573 \\
$S(2s^{2}3p^{4}\;^{3}P\rightarrow2s^{1}3p^{5}\;^{3}P)$ &8.2456&8.3271&8.1723&8.0045&7.4533 \\
$Cl^{+}(2s^{2}3p^{4}\;^{3}P\rightarrow2s^{1}3p^{5}\;^{3}P)$ &9.8117&9.8997&9.7273&9.6933&8.9618 \\
$Cl(2s^{2}3p^{5}\;^{2}P\rightarrow2s^{1}3p^{6}\;^{2}S)$ &9.7143&9.8171&9.6484&9.5153&8.8686 \\
$Ar^{+}(2s^{2}3p^{5}\;^{2}P\rightarrow2s^{1}3p^{6}\;^{2}S)$ &11.3926&11.5061&11.3214&11.1982&10.4901 \\
\hline
Average error &  & 1.00\%& 0.81\% & 2.28\%& 
\end{tabular}
\end{table}

\begin{table}
\caption{Electron transition energy when the upper state is not the lowest
energy multiplet.}
\vspace{0.2in}
\begin{tabular}{lccccc}
\hline
atoms/ions & $\Delta$$E_{HF}$ & $\Delta$$E^{a}$ & $\Delta$$E^{b}$ & $\Delta$$E^{c}$ &
$\Delta$$E_{TDDFT}$  \\
\hline
$B(2s^{2}2p^{1}\;^{2}P\rightarrow2s^{1}2p^{2}\;^{2}D)$ &0.2172&0.2061&0.2696&0.2458&0.2168 \\
$C^{+}(2s^{2}2p^{1}\;^{2}P\rightarrow2s^{1}2p^{2}\;^{2}D)$ &0.3290&0.3216&0.3889&0.3225&0.3325 \\
$C(2s^{2}2p^{2}\;^{3}P\rightarrow2s^{1}2p^{3}\;^{3}D)$ &0.2942&0.2967&0.3755&0.2967&0.3090 \\
$N^{+}(2s^{2}2p^{2}\;^{3}P\rightarrow2s^{1}2p^{3}\;^{3}D)$ &0.4140&0.4305&0.5093&0.4377&0.4433 \\
$Si^{+}(3s^{2}3p^{1}\;^{2}P\rightarrow3s^{1}3p^{2}\;^{2}D)$ &0.2743&0.2799&0.3098&0.2607&0.2864 \\
$Si(3s^{2}3p^{2}\;^{3}P\rightarrow3s^{1}3p^{3}\;^{3}D)$ &0.2343&0.2442&0.2661&0.2445&0.2567 \\
\hline
Average error &  & 3.10\%& 19.67\% & 5.40\%& 
\end{tabular}
\end{table}

\begin{table}
\caption{Electron transition energy when an `s' electron is transferred to
a `d' orbital.}
\vspace{0.2in}
\begin{tabular}{lccccc}
\hline
atoms/ions & $\Delta$$E_{HF}$ & $\Delta$$E^{a}$ & $\Delta$$E^{b}$ & $\Delta$$E^{c}$ &
$\Delta$$E_{TDDFT}$  \\
\hline
$Sc(3s^{2}3d^{1}\;^{2}D\rightarrow3s^{1}3d^{2}\;^{2}G)$ &2.1562&2.1223&2.1800&2.0185&1.8649 \\
$Ti(3s^{2}3d^{2}\;^{3}F\rightarrow3s^{1}3d^{3}\;^{5}F)$ &2.2453&2.2061&2.2963 &2.0242&-----  \\
$Ti(3s^{2}3d^{2}\;^{3}F\rightarrow3s^{1}3d^{3}\;^{3}H)$ &2.3861&2.3649 &2.4190&2.2331&2.0951  \\
$V(3s^{2}3d^{3}\;^{4}F\rightarrow3s^{1}3d^{4}\;^{4}H)$ &2.6098&2.6106 &2.6611&2.4524&2.3266 \\
$Mn(3s^{2}3d^{5}\;^{6}S\rightarrow3s^{1}3d^{6}\;^{6}D)$ &3.1331&3.1199 &3.1637&2.9137&2.8062  \\
$Fe(3s^{2}3d^{6}\;^{5}D\rightarrow3s^{1}3d^{7}\;^{5}F)$ &3.4187&3.4527 &3.4792&3.2326&3.0755  \\
$Co(3s^{2}3d^{7}\;^{4}F\rightarrow3s^{1}3d^{8}\;^{4}F)$ &3.7623&3.7955 &3.7952&3.5489&3.3516  \\
$Ni(3s^{2}3d^{8}\;^{3}F\rightarrow3s^{1}3d^{9}\;^{3}D)$ &4.1204&3.4176 &4.1135&3.8639&3.6351 \\
\hline
Average error & & 2.94\% & 1.33\% & 6.60\%& 
\end{tabular}
\end{table}

\begin{table}
\caption{Electron transition energy when a `p' electron is transferred to
a `d' orbital.}
\vspace{0.2in}
\begin{tabular}{lccccc}
\hline
atoms/ions & $\Delta$$E_{HF}$ & $\Delta$$E^{a}$ & $\Delta$$E^{b}$ & $\Delta$$E^{c}$ &
$\Delta$$E_{TDDFT}$  \\
\hline
$Sc(3p^{6}3d^{1}\;^{2}D\rightarrow3p^{5}3d^{2}\;^{2}H)$ &1.1295&1.2458 &1.4925&1.1366&1.2128  \\
$Ti(3p^{6}3d^{2}\;^{3}F\rightarrow3p^{5}3d^{3}\;^{3}I)$ &1.2698&1.2728 &1.6617&1.2811&1.3586  \\
$V(3p^{6}3d^{3}\;^{4}F\rightarrow3p^{5}3d^{4}\;^{4}I)$ &1.4153&1.4227 &1.8318&1.4297&1.5042  \\
$Mn(3p^{6}3d^{5}\;^{6}S\rightarrow3p^{5}3d^{6}\;^{6}F)$ &1.7270&1.6726 &2.1824&1.7405&1.8073  \\
$Fe(3p^{6}3d^{6}\;^{5}D\rightarrow3p^{5}3d^{7}\;^{5}G)$ &1.8785&2.0061 &2.4249&2.0030&1.9898  \\
$Co(3p^{6}3d^{7}\;^{4}F\rightarrow3p^{5}3d^{8}\;^{4}G)$ &2.1178&2.2778 &2.6522&2.2910&2.1755  \\
$Ni(3p^{6}3d^{8}\;^{3}F\rightarrow3p^{5}3d^{9}\;^{3}F)$ &2.4232&2.5518 &2.8756&2.5923&2.3656  \\
\hline
Average error &  & 4.84\%& 24.33\% & 3.59\%& 
\end{tabular}
\end{table}

\begin{table}
\caption{Excitation energies of some atoms when two electrons are excited.}
\vspace{0.2in}
\begin{tabular}{lcccc}
\hline
atoms/ions & $\Delta$$E_{HF}$ & $\Delta$$E^{a}$ & $\Delta$$E^{b}$ & $\Delta$$E^{c}$ 
  \\
\hline
$Be(2s^{2}\;^{1}S\rightarrow2p^{2}\;^{1}D)$ &0.2718&0.2665&0.3598&0.3108 \\
$B(2s^{2}2p^{1}\;^{2}P\rightarrow2p^{3}\;^{2}D)$ &0.4698&0.4798&0.5779&0.5166 \\
$C^{+}(2s^{2}2p^{1}\;^{2}P\rightarrow2p^{3}\;^{2}D)$ &0.6966&0.7180&0.8122&0.6836 \\
$C(2s^{2}2p^{2}\;^{3}P\rightarrow2p^{4}\;^{3}P)$ &0.7427&0.7312&0.8131&0.7503 \\
$N^{+}(2s^{2}2p^{2}\;^{3}P\rightarrow2p^{4}\;^{3}P)$ &1.0234&1.0143&1.0754&0.9414 \\
$N(2s^{2}2p^{3}\;^{4}S\rightarrow2p^{5}\;^{2}P)$ &1.1789&1.1785&1.2371&1.1668 \\
$O^{+}(2s^{2}2p^{3}\;^{4}S\rightarrow2p^{5}\;^{2}P)$ &1.5444&1.5480&1.5621&1.4178 \\
$O(2s^{2}2p^{4}\;^{3}P\rightarrow2p^{6}\;^{1}S)$ &1.5032&1.4736&1.4180&1.3690 \\
$F^{+}(2s^{2}2p^{4}\;^{3}P\rightarrow2p^{6}\;^{1}S)$ &1.8983&1.8494&1.8129&1.6715 \\
$Mg(3s^{2}\;^{1}S\rightarrow3p^{2}\;^{1}D)$ &0.2578&0.2555&0.2651&0.2612 \\
$S(3s^{2}3p^{4}\;^{3}P\rightarrow3p^{6}\;^{1}S)$ &1.0273&1.0266&1.1306&0.9783 \\
$P(3s^{2}3p^{3}\;^{4}S\rightarrow3p^{5}\;^{2}P)$ &0.8539&0.8680&0.9661&0.8477 \\
$Si^{+}(3s^{2}3p^{1}\;^{2}P\rightarrow3p^{3}\;^{2}D)$ &0.5856&0.6230&0.6979&0.5750 \\
$Si(3s^{2}3p^{2}\;^{3}P\rightarrow3p^{4}\;^{3}P)$ &0.5860&0.5986&0.6703&0.5706 \\
$Cl^{+}(3s^{2}3p^{2}\;^{3}P\rightarrow3p^{4}\;^{3}P)$ &1.2535&1.2516&1.3493&1.1067 \\
\hline
Average error & & 1.71\%& 11.64\% & 6.18\%
\end{tabular}
\end{table}

\end{document}